# The Missing Link of XR: Empathy-Driven Reality for XR and Beyond

Yun Suen Pai* The University of Auckland
Tamil Selvan Gunasekaran† The University of Auckland
Jiashuo Cao‡ The University of Auckland
Kunal Gupta§ The University of Auckland
Andreia Valente¶ The University of Auckland
Ken Jen Lee‖ University of Waterloo
Giulia Barbareschi** University of Duisburg-Essen
Kai Lukoff†† Santa Clara University
Lingyuan Li‡‡ Meta
Ruofei Du Google XR Labs
Fannie Liu JPMorganChase
Jennifer Day Activision
Tanner Person Monash University
Johann Wentzel University of Waterloo
Misha Sra University of California, Santa Barbara
Yuhang Zhao University of Wisconsin-Madison
Theophilus Teo Adelaide University
Elisabeth Andre University of Augsburg
Mark Armstrong Keio University Graduate School of Media Design
Mark Billinghurst The University of Auckland
Danielle Lottridge The University of Auckland
Kinga Skierś Keio University Graduate School of Media Design
Anish Kundu Keio University Graduate School of Media Design
Erica Principe Cruz Center for Transformational Play
Takuji Narumi The University of Tokyo
Kouta Minamizawa Keio University Graduate School of Media Design

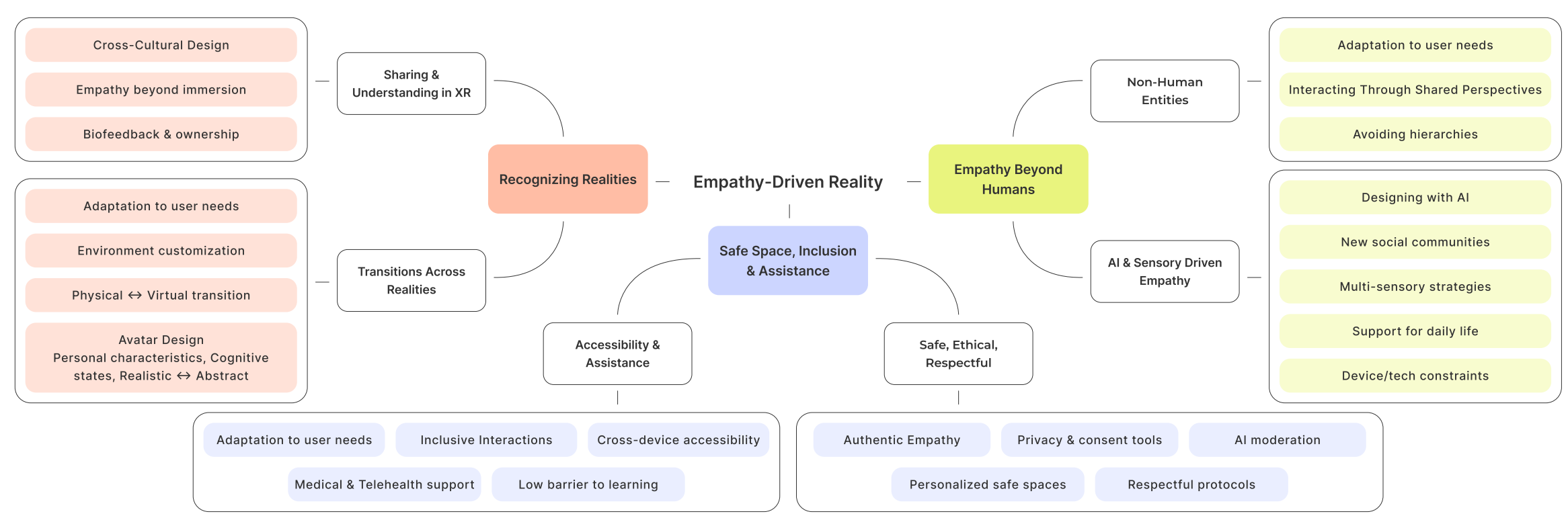

Figure 1: Empathy-Driven Reality (EDR) Design Space.

## Abstract

Extended reality (XR) for socialising is becoming increasingly popular. However, unlike conventional social platforms, XR prioritises embodiment and immersion, factors that strongly impact one's physical and mental states. We envision a future for XR where all users, regardless of abilities and backgrounds, can understand one another, participate, and find safe socialisation spaces. An Empathy-Driven Reality (EDR) is a space where understanding each other's emotional, physical, and cognitive states takes centre stage. It has the potential to enhance empathy beyond how we normally perceive it. To explore this concept, we conducted a hybrid-style workshop over two months with 27 industry and academic researchers in XR, emotion, physiology, assistive technology, and social science. This paper reports on the findings and aims to establish a structure and reference for the 1) design guidelines, 2) research challenges, and 3) potential applications for the future of XR as an EDR. We frame EDR as a conceptual design framework rather than a validated system: a structured design space that links requirements to design mechanisms, together with a research agenda for empathy-centred XR.

*e-mail: yun.suen.pai@auckland.ac.nz
†e-mail: tg469@aucklanduni.ac.nz
‡e-mail: jcao403@aucklanduni.ac.nz
§e-mail: kunal.gupta@auckland.ac.nz
¶e-mail: andreia.valente@auckland.ac.nz
‖e-mail: kenjen.lee@uwaterloo.ca
**e-mail: giulia.barbareschi@uni-due.de
††e-mail: klukoff@scu.edu
‡‡e-mail: lingyuanli9936@gmail.com
e-mail: me@duruofei.com
e-mail: fannie.liu@jpmchase.com
e-mail: jenaday13@gmail.com
e-mail: tanner.person@monash.edu
e-mail: jdwentze@uwaterloo.ca
e-mail: sra@ucsb.edu
e-mail: yuhang.zhao@cs.wisc.edu
e-mail: theo.teo@adelaide.edu.au
e-mail: andre@informatik.uni-augsburg.de
e-mail: mark@keio.jp
e-mail: mark.billinghurst@auckland.ac.nz
e-mail: d.lottridge@auckland.ac.nz
e-mail: kinga.skiers@gmail.com
e-mail: anish@kmd.keio.ac.jp
e-mail: ecruz@cs.cmu.edu
e-mail: narumi@cyber.t.u-tokyo.ac.jp
e-mail: kouta@kmd.keio.ac.jp

# 1 Introduction

This paper introduces *Empathy-Driven Reality* (EDR), a conceptual design framework for extended reality (XR) that foregrounds empathy across embodiment, interaction, and inclusivity. By a conceptual design framework, we mean a structured articulation of the requirements that empathy imposes on XR together with the design mechanisms that can meet them, and an accompanying research agenda. It is intended to orient and generate future systems rather than to constitute an empirically validated design method or a single deployable system. Rather than framing VR as an "empathy machine," EDR conceptualises empathy (cognitive, affective, and compassionate) as a systematic design principle embedded in avatars, environments, and interactions to promote prosocial, inclusive, and ethical engagement.

Chris Milk's documentary "Clouds over Sidra" [1] highlights VR's empathic promise, yet evidence remains mixed regarding its effectiveness in understanding others' emotions [61]. XR also faces challenges in design, ethics, and social acceptance, particularly in supporting diverse representation [27]. Technological constraints limit empathic expression [44, 80], while balancing universal design with individual customisation remains difficult [106, 12, 59]. The use of physiological sensors further raises concerns around privacy, emotional integrity, and consent [58], and societal readiness for everyday XR interaction remains underexplored [82].

EDR aims not only to integrate XR into lived experience but to deepen emotional understanding across human and non-human perspectives, fostering collective empathy and prosocial outcomes. While related visions exist [78, 73], the broader design space has not yet been systematically examined through expert consensus.

To address this gap, we conducted a two-month hybrid workshop with 27 international experts in XR, emotion science, accessibility, and social computing. From this, we derive five definitions, five challenges, 16 design implications (five environment, six interaction, and five avatar designs), and four potentials and applications, organised into three themes: (1) recognising and adapting across realities, (2) redefining safe spaces, and (3) empathy beyond humans.

The contributions of this work are: 1) A definition and framing of EDR as a conceptual design framework for social XR design, distinguished from prior accessibility, privacy, and affective-computing work by treating empathy as the single organising design principle that reframes each of them, 2) a hybrid multi-session expert workshop method for eliciting EDR challenges, guidelines, and applications, and 3) three overarching themes and a structured design space for EDR, represented through a treemap of themes, subthemes, and design implications.

Making empathy the core principle raises questions that do not arise when accessibility, privacy, or affective computing are treated separately, and these questions define the core of EDR. First, how should an empathic signal change as a user moves between XR forms, when accessibility calls for adapting it, and privacy calls for withholding it? Second, when does personalising a system for one user break the shared world that mutual empathy depends on? Third, how should consent and legibility work when the party whose state is conveyed is not human? We return to these questions throughout the paper and use them to focus the design space rather than cover all of it.

# 2 Method

## 2.1 Recruitment and Contributors

Contributors were selected for expertise in XR and HCI and recruited for diversity in disciplinary background and career stage. They were geographically distributed and had an average of 8.1 years of experience (Table 1); all had published peer-reviewed XR research. Primary expertise denotes core disciplinary grounding, while secondary expertise reflects adjacent active interests, enabling analysis shaped by a broad range of perspectives.

The group spanned senior academics, junior academics, and senior industry practitioners, with strong representation in affective and empathic computing (e.g., P5, P6, F12, P20, O23, O25, O26) and accessibility and inclusion (e.g., P3, F8, F9, P11, F15, O27). Additional expertise covered digital wellbeing, immersive collaboration, embodied informatics, human-AI interaction, and computer-mediated communication, complemented by secondary expertise in social robotics (F12), serious games (P7), embodied augmentation (P13, P19), and affective haptics (O25). While not statistically representative, this pool reflects domains intersecting empathy, XR, and socio-technical interaction.

We adopted an authorship-as-participation model: all contributors who engaged in any session were offered co-authorship, as communicated during recruitment and consent. This approach reflects a commitment to collaborative knowledge production. It situates our study within an emerging strand of first-person, autoethnographic, and reflexive design-research methods in HCI, in which researchers draw on their own situated expertise as data [42, 15], alongside the expert-driven synthesis efforts we build on in Section 2.2 [68, 75, 25]. We treat the resulting synthesis as expert perspective rather than neutral observation, and we return to its optimism bias in Section 6. Consent covered recording and documenting all synchronous and asynchronous activities (Zoom and Miro), with access restricted to the research team and contributors. Each contributor also provided at least two related papers, informing the Related Work section.

Initially, 38 researchers agreed to participate; the final cohort comprised 27 contributors (12 senior academics, 11 junior academics, and 4 senior industry practitioners). Attrition was due to logistical constraints rather than conceptual disagreement. Seniority was defined as holding a PhD with at least three years of experience. Six organisers (including the first author) coordinated study design, analysis, and writing; contributors participated in online discussions and reviewed the sections representing their own domains of expertise, so as to safeguard technical accuracy, while the consolidated synthesis and overall framing were circulated to all contributors for comment during member checking and the wrap-up session. No contributor declined authorship or raised concerns about the final framing.

## 2.2 Workshop Organisation

This work builds on prior expert-driven synthesis efforts in HCI [68, 75, 25].

**Session 1: Defining EDR (Online).** A 2-hour session included introductions, a future-sketching exercise (2030–2050), and breakout literature discussions. Moderated groups synthesised visions, identified themes, and created visualisations using Zoom and Miro.

**Session 2: Design Implications (Offline).** Contributors completed a one-week Miro worksheet[2] focused on personas and contexts [30], empathy mapping [26], scenario definition, SWOT analysis [35], and a How Might We design challenge [94] addressing avatar, environment, and interaction design. Summaries of outputs were shared with all participants.

**Session 3: Challenges and Concerns (Offline).** Contributors identified challenges from the persona perspective and categorised them using Self-Determination Theory (autonomy, competence, relatedness) [18], which has been applied to motivation [31] and organisational outcomes [113]. Concerns were prioritised with an Impact–Urgency (Eisenhower) matrix [46], followed by ideation of design strategies.

[1] Clouds Over Sidra: https://www.youtube.com/watch?v=mUosdCQsMkM

[2] EDR Miro Worksheet: https://doi.org/10.17605/OSF.IO/629VX

Table 1: The list of contributors with their specific field of expertise within HCI. Each contributor is unlike typical user study participants; they are also co-authors of this work. All contributors were included as co-authors. O = organising team; F = completed all activity rounds; P = partial participation.

| ID | Gender | Career Stage | Primary Expertise | Secondary Expertise | Year(s) |
|---|---|---|---|---|---|
| F1 | Female | Senior Industry | Cognitive Psychology | Computer-Mediated Communication | ∼ 7 years |
| F2 | Male | Senior Academic | Digital Wellbeing | Social Computing | ∼ 8 years |
| P3 | Female | Junior Academic | Accessibility and Inclusion | Digital Wellbeing | ∼ 2 years |
| F4 | Male | Senior Industry | Interactive Perception | Human-AI Interaction | ∼ 8 years |
| P5 | Male | Junior Academic | Digital Wellbeing | Affective/Empathic Computing | ∼ 3 years |
| P6 | Female | Senior Academic | Human Factor Engineering | Affective/Empathic Computing | ∼ 15 years |
| P7 | Female | Junior Academic | Digital Wellbeing | Serious Games | ∼ 4 years |
| F8 | Female | Senior Academic | Accessibility and Inclusion | Affective/Empathic Computing | ∼ 9 years |
| F9 | Male | Junior Academic | Accessibility and Inclusion | Digital Wellbeing | ∼ 3 years |
| P10 | Male | Senior Academic | Multimodal Interfaces | Embodied Informatics | ∼ 15 years |
| P11 | Male | Junior Academic | Accessibility and Inclusion | Digital Wellbeing | ∼ 2 years |
| F12 | Female | Senior Academic | Affective/Empathic Computing | Social Robotics | ∼ 33 years |
| P13 | Male | Senior Academic | Embodied Informatics | Human Augmentation | ∼ 16 years |
| F14 | Female | Senior Academic | Human-AI Interaction | Spatial Computing | ∼ 10 years |
| F15 | Female | Senior Academic | Accessibility and Inclusion | Digital Wellbeing | ∼ 9 years |
| F16 | Male | Senior Academic | Immersive Collaboration | Multimodal Interfaces | ∼ 6 years |
| F17 | Female | Senior Industry | Computer-Mediated Communication | Social Computing | ∼ 9 years |
| F18 | Female | Senior Academic | Computer-Mediated Communication | Social Computing | ∼ 4 years |
| P19 | Male | Junior Academic | Embodied Informatics | Human Augmentation | ∼ 2 years |
| P20 | Male | Senior Academic | Affective/Empathic Computing | Immersive Collaboration | ∼ 28 years |
| P21 | Male | Senior Industry | Immersive Collaboration | Affective/Empathic Computing | ∼ 10 years |
| O22 | Male | Junior Academic | Positive Computing | Digital Wellbeing | ∼ 2 years |
| O23 | Male | Junior Academic | Affective/Empathic Computing | Accessibility and Inclusion | ∼ 3 years |
| O24 | Male | Junior Academic | Human-AI Interaction | Affective/Empathic Computing | ∼ 1 year |
| O25 | Female | Junior Academic | Affective Haptics | Affective/Empathic Computing | ∼ 1 year |
| O26 | Male | Junior Academic | Affective/Empathic Computing | Digital Wellbeing | ∼ 1 year |
| O27 | Male | Senior Academic | Accessibility and Inclusion | Affective/Empathic Computing | ∼ 8 years |

**Session 4: Potentials and Applications (Offline).** Contributors revisited earlier sketches and personas to envision applications of EDR from 2030 to 2050 using a future-oriented journey map, emphasising speculative and emerging uses.

**Wrap-Up Session (Online).** A final 1-hour Zoom session consolidated findings and gathered additional feedback prior to analysis.

### 2.3 Analysis Method

We conducted thematic analysis [14] supported by affinity diagramming [37] in three phases: transcription, coding, and theming. Two authors independently coded an initial subset and resolved discrepancies through discussion, achieving 87% agreement. Given the exploratory nature and heterogeneous data (sketches, worksheets, discussions), we adopted consensus coding rather than full inter-rater statistics. The remaining data were coded through iterative author meetings. Contributors participated in validation through member checking: after each session, coded theme summaries and example excerpts were circulated for feedback and clarification. Since member checking relied on consensus rather than scored agreement, we also recorded points of divergence. Two recurred and are carried into the themes rather than resolved. First, contributors disagreed over whether biosignal sharing should ever be enabled by default. Some regarded a sensible default as necessary for spontaneous empathy, while others regarded any default disclosure as an unacceptable privacy risk. Second, contributors disagreed over whether extending empathy toward non-human entities enriches empathy between people or competes with it for limited attention and care. We treat these tensions as findings in their own right (Themes 1 and 3).

## 3 Related Work

This section is a synthesis of the literature shared by each contributor at the start of the workshop.

### 3.1 Empathy Across Realities

Empathy is distinguished into cognitive, affective, and compassionate forms [79, 81]. Cognitive empathy involves understanding another's perspective [41], while affective empathy entails sharing their emotional state [63]. Empathy can extend beyond humans to artificial agents perceived as conscious, such as GPT-3 [89]. Compassion differs in that it emphasises motivation to alleviate suffering rather than co-feeling alone [97, 20], and has been framed as dependent on empathy [50]. This work focuses on empathy rather than compassion, as it applies beyond distress contexts and aligns with our design goals [107]. Stepanova et al. identified nine technological strategies for fostering empathy, highlighting the importance of shared embodied experience [103].

Empathy is closely related to altruism: the empathy–altruism hypothesis posits empathic emotion as a driver of prosocial motivation [8, 9]. However, empathy remains difficult to operationalise computationally; while peer support seekers report high empathy, text-based models detect lower levels [108]. Empathy also operates at cultural scales, where cultural empathy supports understanding across differences [23], as demonstrated in systems designed to promote culturally grounded interaction [3].

Physiological signals offer access to emotional and cognitive states [7], though interpretation and integration remain challenging [104, 11, 2, 83]. Prior work has used biofeedback to support empathic awareness [110] and avatars to visualise emotion [122]. Systems such as virtual co-embodiment [47], Animo [54], Heightened Empathy [2], the Transcendental Avatar [99], and AR emotion overlays [112] demonstrate bioresponsive emotion sharing. CAEVR integrates EEG, EDA, and HRV to adapt virtual agents and environments, increasing users' positive affect and empathy [33]. Similarly, empathy-enabled XR combines biosensing with AI-driven agents to dynamically respond to users' emotions [19]. These works motivate our workshop focus on representing and sharing emotion across realities.

Empathy in XR is closely tied to presence, defined as perceiving another's consciousness in virtual space [4, 56, 71, 100]. VR experiences can promote prosocial behaviour through heightened presence [77]. However, findings are mixed: VR may enhance cognitive empathy [115, 117] or primarily affective empathy [61]. Ethical concerns remain, including risks of "toxic empathy" and oversimplified emotional understanding [69].

### 3.2 Safe Spaces in XR

Socially aware XR systems depend on social perception, behaviour synthesis, and the acquisition of social behaviours [87]. Virtual environments such as MMORPGs support identity formation and belonging [32], while systems like Divided Presence [92] and Visual Captions [55] adapt participation and communication dynamics. In social VR, avatar self-representation shapes identity through gender, race, and aesthetics [28], and synchronised breathing in JeL fosters intimacy [105]. Mixed Reality has been proposed as a next step for empathy-driven interaction [114], with platforms such as Geollery supporting user-generated, avatar-based exploration [21]. However, multisensory feedback can also influence social behaviour in unintended ways, including reduced cooperation [64, 109].

Accessibility research stresses co-design with people with disabilities rather than simulating impairment [10]. Systems such as Dementia Eyes [93] demonstrate how visual simulation can foster empathy, while SeeingVR enhances access through magnification and recolouring [128]. Additional approaches include AR magnification for reading [102], Protosound for ambient sound awareness [43], Wearable Subtitles for visualising speech [70], and assistive telepresence in the Avatar Robot Café [5]. Despite these advances, VR hardware remains difficult to access for users with mobility impairments [66], and many users prefer avatars that reflect their assistive devices [127].

Inclusivity in XR also extends beyond disability. VR has been shown to foster empathy in children [36], support intergenerational connection through gamification [90], and provide coping mechanisms for bullying [116, 86]. For LGBTQ+ users, VR can offer spaces of validation and belonging [52], while MR has been found to intensify emotional experiences and avatar preference among gay men [49]. For older adults, multisensory VR designs integrating visual, cognitive, and physical functions improve acceptability and cognitive training [53], and ageing effects further motivate inclusive interaction design [121]. For neurodivergent women and non-binary individuals, VR can function as a safe space for anxiety and emotion regulation [98].

### 3.3 Empathy Beyond Humans

Virtual body ownership [57] shows how self-representation shapes empathic responses, while non-humanoid avatars can increase empathy toward animals [48]. Experiences such as *Justin Beaver VR* [96], where a realistic beaver avatar enables users to feel pain when harmed, and farmed pig embodiment [39] demonstrate both the potential and ethical complexity of animal embodiment, which may also risk reinforcing speciesism.

Cultural empathy research [3, 23] shows that virtual agents can support intercultural understanding. This has been extended through empathic AI systems, including empathic mixed-reality agents [16], empathic overlays [112], and emotion-mirroring avatars for inclusive communication [73, 78]. Multimodal feedback combining physiological, behavioural, and contextual cues maximises perceived empathy [85]. Systems such as *Heightened Empathy* [2] and the *Transcendental Avatar* [99] visualise HRV and EDA as biofeedback, while AR overlays [112] and VRdoGraphy [34] adapt visual and auditory attributes using EEG, EDA, and HRV.

Design strategies further shape empathic experience. Skeuomorphic visuals [51] support intuitive emotion representation, while gaze- and gesture-based interfaces [45] reduce task load and increase co-presence. Abstract environments such as *Isles of Emotions* [91] use spatial metaphors to convey affect and relatedness. Perspective also matters: first-person viewpoints elicit stronger empathy and emotional engagement than third-person views without changing moral decisions [124].

At the same time, caution is required. Prior work warns against toxic empathy [69], as well as surveillance and misuse of empathic systems [22, 74]. Decentralised approaches such as blockchain have been proposed as safeguards [22]. Recent findings on retrospective embodiment further show that who and how one embodies can reshape relational understanding, underscoring the importance of careful empathic role design [125].

## 4 Overarching Themes for EDR

We organise the findings into three themes with sub-themes and design implications. The themes are overlapping lenses on one design space, not mutually exclusive categories. Because empathy cuts across recognition, safety, and non-human relations, concerns such as accessibility and affective disclosure recur across themes by design, each examined from a different angle rather than redundantly.

### 4.1 Theme 1: Recognising and Adapting Across Realities

Theme 1 highlights a core design challenge of EDR: empathy must remain meaningful as users move across physical, virtual, and hybrid spaces. Rather than treating these spaces independently, contributors framed empathy as something that must persist across realities. This introduces a design space in which systems must recognise emotional states, represent them meaningfully, and adapt interaction accordingly while balancing competing constraints such as privacy, comfort, and interpretability.

#### 4.1.1 Empathic Expression and Interpretation in XR

A first mechanism concerns how XR systems can support the expression and interpretation of emotional states. Contributors framed empathy not simply as immersion but as the ability for users to recognise and respond to each other's emotional experiences (P20, F2, F9, P7, F4, F8, P10, F1). EDR functions as an infrastructure for emotional communication by combining XR interaction, physiological sensing, and emotion recognition to enable emotional signals to be expressed, shared, and interpreted across realities (P20, O22, P19).

Designing such empathic channels introduces a tension between *affective disclosure* and *privacy*. Systems capable of detecting and visualising emotional signals can strengthen interpersonal understanding, yet the same mechanisms risk exposing sensitive physiological or emotional information. Contributors therefore emphasised the need for user control over how emotional states are represented or shared.

Several mechanisms were discussed for mediating emotional expression in XR. Embodied interaction allows gestures and body movement to convey affective intent [85], while shared goals can foster social bonding and cooperation [32, 67]. Biofeedback systems can visualise physiological signals such as heart rate to make internal emotional states perceptible [54], while emotion mirroring through facial expressions or vocal tone can further strengthen mutual awareness [105, 24].

Avatar design also plays a crucial role. Prior work on the Proteus Effect suggests that avatar characteristics can influence both self-perception and social interaction [123]. However, the design of emotionally expressive avatars introduces another tension: increasing expressive fidelity can enhance empathic interpretation, yet highly realistic avatars may raise concerns around identity exposure, surveillance, or social pressure.

These findings suggest that EDR systems must treat empathic communication as a configurable channel rather than a fixed feature.

#### 4.1.2 Maintaining Empathy Across Reality Transitions

A second challenge concerns how empathic understanding can persist as users transition between physical and virtual environments. Contributors described the difficulty of maintaining emotional continuity across realities, highlighting issues such as physical discomfort from head-mounted displays (F15), motion sickness, and the difficulty of transferring virtual experiences into real-world behaviour (F12).

This raises a tension between *immersion* and *accessibility*. Highly immersive XR environments may strengthen emotional engagement, yet they can introduce physical or cognitive barriers that limit who can comfortably participate. Allowing users to modify visual, auditory, or atmospheric features, such as colour, lighting, sound, or ambient mood, enables environments to adapt to individual preferences and sensory needs (F1, F9, F17). However, personalisation introduces a competing tension between *individual comfort* and *shared reality*: if each user experiences a heavily customised environment, maintaining a coherent shared experience becomes more difficult.

A related mechanism involves adaptive environments that respond to users' physical or emotional states. Examples include remapping input interactions to accommodate different abilities (F9, F14) or environments that respond to emotional signals by adjusting atmosphere or interaction flow (F15-F17). While adaptive environments can increase accessibility, they introduce a tension between *system autonomy* and *user agency*, as users may feel uncomfortable if systems respond to emotional signals in ways they cannot control.

Another key challenge concerns transitions along the reality-virtuality continuum [65]. Physical objects and assistive technologies could transfer into XR spaces [118]. Such continuity may strengthen users' sense of ownership and connection between environments (F1, F16, F17). However, maintaining continuity introduces trade-offs between *physical fidelity* and *design flexibility*, since faithfully replicating real-world environments may limit the expressive possibilities of virtual spaces.

Empathic communication occurs differently in intimate one-to-one interactions compared to larger group settings. Systems may therefore need to adapt visual cues or interaction mechanisms depending on group size, helping users perceive collective dynamics such as group attention or emotional tone [2]. Yet designing such feedback introduces another tension between *social awareness* and *information overload*, as excessive emotional signalling may overwhelm users rather than support understanding.

Participants envisioned avatars that could represent personal characteristics such as abilities or social roles, as well as avatars capable of visualising otherwise invisible emotional or physiological states. However, the degree of avatar realism remains context dependent. Photorealistic avatars may support subtle emotional interpretation [40], while more abstract representations can amplify emotional states or avoid stereotyping [2]. This highlights a final design tension between *realism* and *interpretability*: realistic avatars may improve fidelity but risk social pressure or bias, whereas stylised avatars can communicate emotion more clearly but reduce personal authenticity.

Taken together, Theme 1 reveals that recognising and adapting across realities is not a single design problem but a set of interconnected trade-offs.

### 4.2 Theme 2: Redefining Safe Space Towards Inclusion, Accessibility and Assistance

Theme 2 explores how EDR environments can function as inclusive social infrastructures. Contributors emphasised that accessibility and assistance should not be treated as optional features but embedded into the core design of XR platforms. Safe spaces in this context are not merely protected environments but systems that actively support diverse abilities, identities, and communication styles while safeguarding privacy and autonomy.

#### 4.2.1 Designing Accessible Empathic Environments

A central design direction concerns how XR systems can support users with diverse abilities while maintaining a shared social environment. Participants framed accessibility not simply as technical accommodation but as the ability for users with different capabilities, backgrounds, and identities to participate meaningfully in social interaction (F8, F15). EDR systems should follow an ability-based design philosophy that focuses on users' capabilities rather than limitations [120], enabling individuals to augment their abilities and integrate XR technologies with existing assistive tools (F2, F9, F14, F15). Heavily customised interfaces may support participation but risk fragmenting the shared experience.

Several mechanisms were proposed to navigate this tension. One involves accessible environment design, where visual, auditory, and spatial features can be configured to support different sensory needs. For example, spatialised sound cues can assist visually impaired users in navigating virtual environments (F4), while visual symbols or environmental cues may support communities such as LGBTQAI+ users in recognising inclusive spaces (F18). Environments must remain interpretable across users with different configurations, raising challenges for maintaining shared spatial awareness.

Interaction design also emerged as a critical mechanism. Contributors envisioned multisensory communication channels combining haptics, sound, and visual signals to convey social and environmental information. Captioning systems were discussed not only for speech but also for contextual environmental cues, such as footsteps approaching from behind. These extensions support richer emotional awareness, yet they introduce a tension between *informational richness* and *cognitive load*, as excessive sensory feedback may overwhelm users rather than assist them.

Avatar design represents another key site for accessible interaction. Participants described avatars that embed assistive features such as subtitles, translation tools, and virtual sign-language interpreters (F4). Avatars may also represent diverse identities or abilities, enabling marginalised users to be recognised within social environments (F12). However, such representational mechanisms raise a design tension between *visibility and protection*: representing disability or identity can foster empathy but may also expose users to unwanted attention or bias.

Learning barriers were also highlighted as a challenge. XR systems often require unfamiliar interaction techniques that can discourage participation, particularly for older adults or individuals with limited motor control [38]. Contributors suggested that contextual cues, simplified interactions, and alternative input modalities can reduce learning friction (F1, F15). Yet simplifying interactions introduces a trade-off between *ease of entry* and *expressive depth*, as highly simplified interfaces may limit the richness of social expression.

Finally, contributors envisioned EDR environments supporting medical and therapeutic contexts, including telemedicine, rehabilitation, and social recovery spaces (F1, F15). Post-treatment environments may allow individuals to reconnect with social communities and develop empathy through shared experiences (F16). Such applications introduce a tension between clinical reliability and open social interaction, as therapeutic spaces must stay trustworthy while supporting organic interaction.

#### 4.2.2 Governance, Privacy, and Ethical Safety

Beyond accessibility, safe spaces require governance mechanisms that regulate how emotional information and social behaviour are managed within XR environments. Because EDR systems often rely on physiological sensing or emotional inference, participants emphasised the importance of ethical safeguards around how empathic signals are collected and interpreted.

One design challenge concerns the tension between *emotional transparency* and *data privacy*. Systems capable of visualising emotional states may deepen social understanding, yet emotional data are inherently sensitive. Contributors therefore proposed that empathic assessment should combine self-report, behavioural signals, and physiological data while giving users clear control over how it is shared. Safeguards such as strong personal data protection, transparent data policies, and explicit design principles prioritising inclusivity were seen as essential to prevent exploitative or performative empathy.

Contributors also envisioned privacy filters that control what emotional or physiological signals are visible to others (F8). Concepts such as a "privacy avatar" could visually communicate disclosure preferences, similar to visual consent cues in prior work [55]. These preserve authenticity while letting individuals regulate boundaries, but introduce a tension between authentic expression and strategic self-presentation, as users may selectively conceal emotional signals.

Respectful interaction also requires safeguards against harassment or harmful behaviour. Contributors suggested a combination of activity-based interaction constraints (F15), content moderation mechanisms (F9), and AI-assisted detection of harassment or hate speech (F18, F14). AI moderation may help protect vulnerable communities, yet it introduces another tension between *safety* and *overregulation*. Since moderation tools can buffer harassment but may exclude users if applied too aggressively [29]; contributors therefore advocated hybrid human-AI governance systems that incorporate community oversight and appeal mechanisms.

Finally, participants envisioned EDR environments supporting personal control over social exposure. Users may create spatial "bubbles" that regulate who can approach or communicate with them (F9), allowing individuals to temporarily withdraw from overwhelming interactions. Because many users avoid voice communication due to harassment or anxiety, systems should also support voice-optional participation through rich captioning, expressive emotes, and consent-based turn-taking protocols [17].

Taken together, Theme 2 highlights that safe spaces in Empathy-Driven Reality are not static environments but dynamic infrastructures balancing accessibility, representation, privacy, and governance.

### 4.3 Theme 3: Empathy Beyond Humans

The third design space emerging from our workshops concerns empathy directed toward specific non-human entities, namely AI agents, animals, and ecological systems, rather than more-than-human interaction as an entire field of theory. Contributors explored how XR systems such as EDR might translate internal states from AI agents, animals, or environmental systems into perceivable forms that humans can interpret and emotionally respond to. Rather than representing isolated applications, these discussions revealed a broader design direction: XR enables empathy across ontological boundaries by making otherwise inaccessible perspectives interpretable through representation, simulation, and sensory translation. In this sense, EDR functions as an empathic mediator between humans and non-human entities. We frame this as an opening onto more-than-human and multispecies HCI rather than a retreat from it: EDR contributes XR-specific mechanisms, perspective translation, embodied non-human viewpoints, and environmental cues, while remaining cautious about anthropomorphism and the limits of rendering non-human experience.

However, extending empathy in this way introduces a central tension between *interpretability* and *authenticity*. Translating non-human states into human-understandable experiences risks oversimplifying or anthropomorphising complex processes. The following design explorations illustrate how different mechanisms navigate this tension.

#### 4.3.1 Sensory Translation and AI-Mediated Empathy

One direction explored how empathy may emerge through *sensory translation*. Contributors emphasised that empathic experiences in XR often rely on multisensory engagement, where visual, auditory, tactile, or physiological signals reinforce one another to convey affective meaning [64]. For instance, coupling tactile feedback with virtual environments can intensify emotional engagement [111], while multimodal interaction may strengthen users' sense of connection with others or with virtual agents [60].

Physiological sensing and biofeedback emerged as a potential mechanism for translating internal emotional states into perceivable signals. Biosignals may be visualised through avatar changes, environmental responses, or shared sensory cues that allow others to perceive emotional states that would otherwise remain hidden [84, 112]. Such mechanisms build on prior work showing that immersion and embodiment can amplify empathic engagement in XR [72, 119].

However, increasing the visibility of internal states introduces a trade-off between *affective transparency* and *interpretation accuracy*. Physiological signals are often noisy or ambiguous, and their interpretation can vary across contexts and individuals. Overreliance on automated emotional inference may therefore misrepresent users' experiences or create misleading empathic cues. AI was discussed as both an enabler and a challenge. Intelligent agents may support empathy by interpreting sensory data, adapting interactions, or assisting users with limited mobility. Yet, contributors raised concerns about the reliability and latency of AI-driven emotion recognition, particularly when interpreting complex affective or social cues. If AI agents incorrectly infer emotional states or behave inconsistently with human expectations, they may undermine rather than strengthen empathic interaction.

This creates a second tension between *automation and agency*. While AI can enhance accessibility and responsiveness, excessive automation risks reducing users' control over how their emotions or behaviours are represented. Practical constraints also shape this design space. Current XR systems rely on multiple external sensors, additional power sources, and wearable attachments, which may reduce comfort and usability [126]. These constraints suggest that achieving seamless empathic sensing may require new hardware paradigms that integrate physiological sensing directly into XR devices.

#### 4.3.2 Empathy Towards Non-Human Entities

A second direction explored how empathy might extend beyond human or AI agents toward animals, environments, and ecological systems. Contributors proposed representing such entities through avatars or environmental visualisations that communicate their internal states or conditions.

For example, AR overlays could visualise ecological health indicators within physical environments, enabling users to perceive environmental changes that would otherwise remain invisible. A lake's ecological condition might be represented through colour, motion, or atmospheric changes, conveying environmental wellbeing in affective terms. Prior work suggests that such representations can foster empathy for wildlife and ecological systems by translating complex environmental processes into perceptible experiences [76, 96, 95].

Shared AR environments could allow multiple users to perceive and discuss environmental states together, supporting collaborative reflection and decision-making and collective sensemaking about non-human systems. Yet, this design direction raises another tension between *engagement and fidelity*. Representing environmental or animal states through simplified visual metaphors may increase accessibility and emotional impact, but such representations risk anthropomorphising or oversimplifying complex ecological processes.

Across these explorations, the defining mechanism of Theme 3 is *perspective translation*. XR systems such as EDR can mediate encounters between humans and non-human entities by translating unfamiliar sensory logics or internal states into experiences that users can perceive and interpret. The design challenge lies in balancing interpretability with authenticity: enabling meaningful empathic engagement without collapsing the diversity of non-human forms of experience.

## 5 Discussion

To ground the discussion, we present a consolidated design space of Empathy-Driven Reality (EDR) in Figure 1. Our design space operates primarily at a conceptual-theoretical level. This approach aligns with prior XR work that articulates design commitments before full empirical validation [1]. Similarly, EDR synthesises insights from expert discourse and prior literature and should be understood as a foundational scaffold upon which future empirical systems and evaluations can be built.

### 5.1 Key Conceptual Advances of EDR

Our thematic analysis across empathy research, XR design, affective computing, and socio-technical interaction reveals several advances that extend current thinking.

A key insight from Theme 1 is that empathic meaning does not remain stable across physical, augmented, and virtual contexts. Many XR systems derive affective cues directly from sensor data, framing empathy primarily as a recognition or classification problem [104, 84]. However, emotion theory shows that affective states are constructed and context-dependent rather than universal signals [6, 7]. EDR therefore argues that XR systems should support cross-reality continuity, allowing empathic cues to persist, decay, or transform as users traverse the reality-virtuality continuum [65]. Rather than collapsing empathy into singular metrics, EDR treats empathic interpretation as a multi-sensory and socially co-constructed process [105, 107]. This perspective also echoes critiques cautioning against simplistic narratives about empathy effects in VR [69, 119]. The conceptual advance is thus that **empathic XR requires infrastructures that preserve contextual richness and relational meaning across realities**, rather than simply detecting or displaying affect. Preserving contextual richness implies that an empathic cue should carry its provenance and confidence with it, so that a signal generated in a VR encounter is re-situated, rather than re-detected from scratch, when the same users later meet in AR or in the physical world. Empathic state thus becomes something the system helps users carry and reinterpret across a transition, not a per-frame classification recomputed at each boundary. This reframes the core design problem from momentary emotion recognition to the continuity and accountable representation of affect over time.

Theme 2 advances this view by positioning accessibility, safety, and assistance as infrastructural conditions for empathic interaction. Accessibility constraints, social asymmetries, and harassment risks in social VR can shape whether empathic engagement is possible [66, 88, 52]. Ability-based design further argues that systems should adapt to users rather than requiring users to adapt to systems [120]. Contributors therefore emphasised architectures that enable negotiated disclosure of affective information [51], embed adaptive support for diverse sensory and cognitive needs [128], and protect vulnerable groups through technical and social mechanisms [100, 58]. In this view, "safe space" is not a static property but an ongoing relational achievement. EDR thus **unifies accessibility, inclusion, and empathic intent by positioning empathy as a guiding design principle for system rules and affordances**, expanding current narratives around empathic interfaces toward empathic infrastructures.

Theme 3 extends the discourse by proposing that empathy in XR may involve interactions with AI agents, animals, and environmental systems. Empathic AI raises questions about affective attunement, transparency, and social acceptability, echoing work on empathic agents and socially aware interfaces [87, 3, 89]. Ecological or animal-focused empathy requires representations that avoid simplistic anthropomorphism while still enabling perspective-taking [76, 95, 96]. Similarly, non-human embodiment raises questions of sensory mapping, intentionality, and ethics that extend debates on virtual body ownership [62, 101]. EDR therefore reframes **more-than-human empathy as an ecosystem of relational design possibilities**, each requiring tailored representational strategies, governance models, and evaluation methods.

### 5.2 Applicability of the EDR Design Space to Existing XR Practices

Table 2 maps each EDR theme to its conceptual requirements and potential design mechanisms. These mechanisms, confidence annotations, multi-layer empathic cues, disclosure control panels, contextual gating, AI-mediated inference explanations, and non-human perspective-taking modes, are the concrete levers through which each theme's requirements can be met. The scenarios instantiate specific subsets of them. To illustrate applicability, we ground these principles in three XR scenarios. The scenarios are author-synthesised analytical projections rather than validated deployments. Each was constructed from the personas and contexts developed in Session 2 and the future journey maps produced in Session 4, then reworked by the organising team into an illustration. They show how EDR concepts may inform practice, and each names the Table 2 mechanisms it instantiates so that the path from requirement to mechanism to scenario is explicit.

One scenario concerns social VR ecosystems aligned with Theme 1: Recognising and Adapting Across Realities. Platforms such as VRChat[3], Meta Horizon Worlds[4], and RecRoom[5] already support diverse avatar identities and expressive interactions, yet users often lack cues to interpret the affective states of unfamiliar others. EDR could introduce consensual affective disclosure through avatar-adjacent visual signals, such as colour modulation, particle motion, or breathing animations, derived from user-controlled biosignal trends. Rather than exposing raw physiological data or inferring discrete emotions, these cues would communicate only stable patterns (e.g., rising arousal), consistent with research cautioning against reductive emotion mapping [6, 7, 85]. Since social VR platforms already employ safety layers and trust systems [54, 105], EDR could integrate empathic cues within these existing privacy and safety infrastructures. This scenario instantiates the Recognising & Adapting mechanisms in Table 2: multi-layer empathic cues, cross-reality continuity indicators, and confidence annotations attached to the stable patterns being disclosed.

A second scenario involves cross-reality collaboration where users interact across VR, AR, and physical environments. This context reflects Theme 2: Redefining Safe Spaces. Prior work highlights the importance of transparency, consent, and user-controlled communication in mediated environments [55, 120]. EDR could support a "Disclosure Control Panel" allowing users to configure what empathic cues are shared across realities. These settings could be visualised through small emblems attached to avatars, signalling disclosure preferences. This approach echoes existing work on privacy boundaries and safe participation in social VR [17, 29, 67]. Time-limited sharing and explicit opt-in mechanisms also align with research addressing harassment and vulnerability in immersive environments [21, 52, 88]. It instantiates the Redefining Safe Space mechanisms: disclosure control panels, selective visibility layers, and contextual gating with opt-out affordances.

[3]VRChat: https://hello.vrchat.com/

[4]Meta Horizon Worlds: https://horizon.meta.com/

[5]RecRoom: https://recroom.com/

Table 2: Constituent components of the EDR conceptual design space, mapping each theme to its underlying conceptual requirements and the design mechanisms that operationalise them.

| EDR Theme | Conceptual Requirements | Design Mechanisms / Levers |
|---|---|---|
| **Recognising & Adapting Across Realities** | *Interpretive Ambiguity & Confidence Signalling:* Empathic inferences should acknowledge uncertainty rather than assert emotional truth. *Ability-Based Empathic Interpretation:* Empathic cues must adapt to sensory, cognitive, and cultural differences. *Cross-Reality Continuity:* Empathic signals should persist, decay, or transform across XR modalities without collapsing context boundaries. | Confidence annotations and uncertainty tags; multi-layer empathic cues (visual, haptic, behavioural); adaptive empathic calibration profiles; cross-reality continuity indicators. |
| **Redefining Safe Space in XR** | *Consent-First, Granular Affective Disclosure:* Users control when, how, and to whom empathic information is shared. *Negotiated Emotional Legibility:* Empathic transparency should be co-determined in social or professional contexts. *Protection from Emotional Overexposure:* Empathy should not be forced or weaponised. | Disclosure control panels; selective visibility layers; contextual gating of empathic cues; interruptibility and opt-out affordances. |
| **Empathy Beyond Humans** | *Multi-Directional Empathic Relations:* Empathy may extend to AI agents, non-human entities, or ecological systems. *Non-Normative & Disruptive Empathic Experiences:* Empathy may involve discomfort, frustration, tension, or collective grief. *Translation Layers for Non-Human Signals:* Empathic exchange requires interpreting heterogeneous sources. | AI-mediated empathic inference explanations; embodied non-human perspective-taking modes; environmental or agent-based empathic cues; multi-species or non-human storytelling lenses. |

A third scenario concerns AI-driven agents that interpret or respond to human affective cues, directly engaging Theme 3: Empathy Beyond Humans. XR research shows that computational agents can influence emotion, trust, and relational perception [63, 126]. Explainable affective interfaces also highlight the need for transparency when interpreting emotional signals [6, 7, 85]. Under EDR, AI agents might adapt vocal tone, pacing, or spatial behaviour based on biosignal trends while explicitly signalling interpretive uncertainty. For example, subtle avatar animations or explanatory overlays could indicate probabilistic reasoning rather than definitive emotional judgments. Such approaches extend embodied agent research in affective computing [2, 50] while avoiding overclaiming about emotional understanding. Instead of "reading emotions," agents become reflective partners that foreground interpretive ambiguity, aligning with ongoing discussions about anthropomorphism and AI-mediated empathy [11, 105, 54]. It instantiates the Empathy Beyond Humans mechanisms: AI-mediated empathic inference explanations and, where an agent stands in for a non-human entity, embodied non-human perspective-taking modes.

Together, these scenarios demonstrate how EDR principles can integrate with established XR design workflows. Avatar augmentation, disclosure interfaces, and environmental biofeedback visualisations are already common XR components; EDR reframes them through an explicitly empathic lens, making the design space actionable for practitioners.

### 5.3 Recognising Realities: Embodiment and Measurement

A further challenge is how empathy should be defined and measured across XR contexts, where embodied avatars, biofeedback, and adaptive environments blur the boundaries between cognitive, affective, and compassionate empathy [79, 81]. For EDR, the research agenda lies in integrating physiological, behavioural, and self-report signals without collapsing empathy into a single metric [108], and in longitudinal study of how empathic states transition across realities [62, 101].

### 5.4 Safe Spaces and Ability-Based Design

The workshop highlighted that empathy in XR is inseparable from accessibility and inclusion. Safe spaces must therefore function as infrastructures enabling diverse participation rather than simply harm-free environments. What is specific to EDR here is not that XR should be accessible or safe, which is true of XR broadly, but the direction of the dependency. Because empathy is the organising goal, accessibility becomes a precondition for empathic participation rather than a general usability property, and privacy becomes control over emotional and physiological disclosure rather than data protection in the abstract. Safety and inclusion enter the framework as conditions on whether empathic exchange is possible at all, which is precisely why they recur across themes. This perspective aligns with ability-based design, which emphasises adapting systems to individual capabilities [120]. Accessibility research also stresses the importance of authentic participation rather than simulated disability experiences [10]. Avatar representations that reflect assistive technologies or marginalised identities can further support representation and empathic understanding [5, 127].

However, tools and policies often struggle to keep pace with emerging social VR risks. AI-supported moderation is increasingly explored to address toxicity at scale [88]. Studies suggest that hybrid moderation combining human judgment with AI support may offer the most balanced approach. These insights inform the EDR design space: empathy-driven XR must embed proactive safeguards, including accessibility features, identity protection, community norms, and hybrid moderation infrastructures [88]. Such systems aim to empower users, particularly those from marginalised groups, to participate safely in cross-reality interactions.

Safe spaces also require strong privacy and user control. Systems such as Divided Presence [92] and Visual Captions [55] demonstrate how adaptive participation and multimodal support can reduce exclusion. Building on this work, EDR environments should allow users to configure sensory and interaction boundaries, from minimal environments that reduce overload [85] to multimodal notifications that support accessibility [43, 70]. Evaluating these adaptive environments over time will be crucial for establishing safe spaces as a foundation for empathic XR.

### 5.5 Empathy Beyond Humans: Toward Posthumanist EDR

The workshop also explored extending empathy beyond human-to-human relations. Studies show that non-human avatar embodiment can foster empathy toward animals [39, 48, 96], while AR overlays can visualise environmental states such as ecosystem health [95]. Empathic AI agents and multimodal feedback systems further demonstrate how non-human entities can participate in affective interaction [16, 85, 112].

These directions resonate with posthumanist HCI, which critiques strictly human-centred design and explores multispecies or ecological perspectives. We read EDR's translation layers, multispecies lenses, and non-human storytelling modes (see Table 2) as concrete XR openings onto this programme rather than as substitutes for its theory. However, such approaches carry risks. Animal embodiment may unintentionally reinforce species hierarchies [39], while empathic AI can reproduce misleading or manipulative behaviours [69]. Future research must therefore develop critical design methods that clarify both the possibilities and limits of empathising with non-human entities.

### 5.6 Risks, Misuse, and Provocative Potentials of EDR

Physiological sensing within EDR introduces risks of emotional surveillance and misuse [22, 74]. Empathic design may also produce what Nakamura terms "toxic empathy," where emotional exposure overwhelms or manipulates users rather than supporting them [69]. Additionally, XR infrastructures may carry environmental costs due to high computational demands [13]. These risks highlight the need for ethics to be embedded directly within system design.

Workshop participants proposed strategies including privacy filters and "privacy avatars" that visualise user-controlled data sharing [22]. Consent mechanisms must also address the sharing of emotional states by providing transparency and granular control. We therefore propose three ethical principles for EDR. First, privacy-by-design ensures that emotional data remains under user control through granular consent and clear feedback about data sharing. Second, sustainability requires energy-aware sensing architectures and lightweight multimodal processing to reduce environmental impact [13]. Third, cultural accountability emphasises co-design with diverse communities, recognising that emotional expression varies across cultural contexts [6]. These principles extend existing work on empathic computing ethics by shifting responsibility toward proactive design practices [22, 74].

EDR also invites exploration of non-normative empathic experiences. Rather than only supporting calm or prosocial interactions, empathic systems may involve shared frustration, collective mourning, or other emotionally intense encounters. Such possibilities resonate with affective and critical design traditions that treat discomfort as reflective rather than purely negative [10, 69]. Emotional intensity in VR does not necessarily produce prosocial outcomes, highlighting the need for careful design assumptions [117]. Biosignal-mediated synchrony may amplify collective arousal [85], while embodiment research shows that empathy can also provoke defensive or adversarial reactions [76].

Whether empathic experiences are supportive, disruptive, or confrontational, their design must prioritise autonomy, consent, cultural interpretation, and emotional safety [10, 69, 84].

## 6 Limitations and Future Work

As noted in Section 2.1, we initially recruited 38 experts, yet participation at that scale proved challenging. A primary reason was participants' busy schedules and competing professional and personal commitments. Timezone differences further complicated coordination. Future iterations could better support large-scale collaboration by segmenting discussions by timezone and supplementing online activities with occasional in-person meetups.

Our use of Miro boards replicated in-person workshop activities in a digital format, but engagement remained lower than expected. Despite the incentive of co-authorship, contributor motivation decreased over time, resulting in incomplete participation. Twelve contributors completed all rounds, while the remainder contributed to at least one activity (excluding core organisers). Future work may improve engagement through strategies such as gamification, stronger onboarding and training, and more extensive pilot testing. Session 2 was also considerably longer than the others, requiring greater sustained concentration and cognitive effort. Future designs should consider shorter and more balanced activities, for example, deploying 30-minute sessions every three days rather than 2-hour sessions per week.

The EDR design space presented in this paper remains primarily theoretical, emerging from expert synthesis and conceptual analysis rather than empirical deployment. Its applicability should therefore be interpreted as prospective and generative. While we provide mappings to existing XR systems, these are analytical projections rather than empirical evaluations. Future work should validate EDR through real-world implementations, examining how its principles influence user experience, safety, and empathic outcomes across diverse XR contexts. In particular, the design mechanisms in Table 2 and the scenarios in Section 5.2 are the natural targets of a subsequent implementation phase, in which the personas and journey maps generated here would be enacted as working prototypes and evaluated with users beyond the authorship group.

Translating biosignals or affective indicators into empathic cues also remains technically and ethically challenging. Current XR platforms rarely support reliable, low-latency, and privacy-preserving affect sensing [84, 104, 126]. Because EDR relies on contextual and consentful representations of internal states, future systems must address these constraints before the paradigm can be widely applied.

Finally, our methodology shares a limitation noted by Mueller et al. [68]: the authorship-as-participation model may introduce optimism bias, as contributors are often invested in advancing the field. Although participants represented diverse XR backgrounds, several perspectives (e.g., cognitive neuroscience, environmental psychology, animal-computer interaction, non-Western design traditions, and lived-experience experts) were underrepresented, limiting the generalisability of the design space. Nevertheless, this approach provides insights from individuals deeply familiar with the state of the art. Future work could incorporate external critical reviewers outside the authorship group, conduct post-hoc member checking with independent XR practitioners, or run dissensus-oriented sessions that explicitly invite critiques and failure cases of empathic XR. Another strategy is mixed-participation sampling, where only a subset of contributors become co-authors while others participate without authorship obligations, reducing pressure toward consensus.

## 7 Conclusion

This paper presents the outcomes of a two-month hybrid workshop involving 27 industry and academic researchers from diverse backgrounds. Their collective insights were synthesised into three overarching themes that frame the definitions, design implications, challenges, and potentials of empathy-centric XR design. We envision EDR as an important consideration for future XR systems. The three themes: 1) recognising and adapting across realities, 2) redefining safe space towards inclusion, accessibility, and assistance, and 3) empathy beyond humans, provide guiding principles for XR design while emphasising the central role of diversity and emotional understanding in shaping future experiences. The contribution of EDR is not a new list of XR concerns but a reorganisation of familiar ones around empathy as the design principle, and its value lies in the tensions between them, which we have named rather than dissolved and which set the agenda for empathy-driven XR.

## Acknowledgments

Generative AI tools were used at the paragraph level to improve readability and style of text originally envisioned and written by the authors, and to adjust figures (e.g., improving clarity, layout, or labelling) based on author-created originals. The AI assistance was limited to copy-editing and visual refinement; all conceptual content, study design, analysis, and interpretation were created by the authors. All AI-assisted text and image adjustments were carefully reviewed, revised, and approved by the human authors, who remain fully responsible for the final manuscript. No confidential or proprietary data was provided to the AI system. We sincerely thank Zikun Chen from ProtoOwl[6] for her assistance with the workshop and figure design. This work is supported by the University of Auckland Faculty of Science Research Development Fund Grant Number 3731533, JST Moonshot R&D Program (JPMJMS2013), and JST COI-NEXT (JPMJPF2203).

[6] ProtoOwl: https://protoowl.com/